\documentclass[seceq]{ptptex}
\usepackage{graphicx}
\usepackage{epsfig}

\title{Heavy quark potential and QCD beta function from a deformed $AdS_5$ model
}

\author{Song He$^{1}$, Mei Huang$^{1,2}$, Qi-Shu Yan$^{3,4}$}

\inst{$^{1}$ Institute of High Energy Physics, Chinese Academy of
Sciences, Beijing, China \\
$^{2}$ Theoretical Physics Center for Science Facilities, Chinese
Academy of Sciences, Beijing, China\\
$^{3}$ Department of Physics, University of Toronto, Toronto,
Canada\\
$^{4}$ College of Physical Sciences, Graduate University of
Chinese Academy of Sciences, Beijing, China}

\abst{We show that in a deformed ${\rm AdS}_5$ model with an
explicit infrared cutoff included in the logarithmic correction
$-c_0\log[(z_{IR}-z)/z_{IR}]$, the heavy quark Cornell potential can
be fitted very well, the corresponding beta-function agrees with the
QCD beta-function at 2-loop level reasonably well, and its dual
dilaton potential is bounded from below in infrared. The results are
compared with those in the Andreev-Zakharov model and the
Pirner-Galow model.}

\begin{document}

\maketitle

\section{Introduction}

The AdS/CFT duality \cite{dual} has been widely used to discuss the
meson spectra and dense and hot quark matter. The string description
of realistic QCD has not been successfully formulated yet. Many
efforts are invested in searching for such a realistic description
by using the "top-down" approach, \textit{i.e.} by deriving
holographic QCD from string theory, as well as by using the
"bottom-up" approach, \textit{i.e.} by examining possible \textit{holographic%
} QCD models from experimental data and lattice results.

In the "bottom-up" approach, the most economic way is to search for
a deformed ${\rm AdS}_5$ metric
\cite{KKSS2006,Andreev:2006ct,Andreev:2006vy,Shock-2006,Ghoroku-Tachibana,
Csaki:2006ji,Gursoy,Zeng:2008sx,Pirner:2009gr,Brodsky:2010ur}, which
can describe the known experimental data and lattice results of QCD,
e.g. hadron spectra and the heavy quark potential. The simplest
holographic QCD model is the hard-wall ${\rm AdS}_5$ model
\cite{Hardwall-Polchinski}, which can describe the lightest meson
spectra in $80-90\%$ agreement with the experimental data. However,
the hard-wall model cannot produce the Regge behavior for higher
excitations. It is regarded that the Regge behavior is related to
the linear confinement. It has been suggested in Ref.
\cite{KKSS2006} that a negative quadratic dilaton term $-z^2$ in the
action is needed to produce the right linear Regge behavior of
$\rho$ mesons or the linear confinement. The most direct physical
quantity related to the confinement is the heavy-quark potential.
The lattice result which is consistent with the so called Cornell
potential \cite{Cornell} has the form of $ V_{Q{\bar
Q}}(R)=-\frac{\kappa}{R}+\sigma_{str}R+V_0$. Where $\kappa\approx
0.48$, $\sigma_{str}\approx 0.183 {\rm GeV}^{2}$ and $V_0=-0.25 {\rm
GeV}$, the first two parameters can be interpreted as
$\frac{4\alpha_s}{3}$ and QCD "string" tension, respectively.

In order to produce linear behavior of heavy flavor potential,
Andreev and Zakharov in Ref.\cite{Andreev:2006ct} suggested a
positive quadratic term modification \cite{Andreev:2006vy} in the
deformed warp factor of the metric, which is different from the
soft-wall model in \cite{KKSS2006}. In Ref. \cite{White:2007tu}, the
authors found that the heavy quark potential from the positive
quadratic model is closer to the Cornell potential than that from
the backreaction model \cite{Shock-2006}, which contains higher
order corrections.

It is clearly seen from the Cornell potential that the Coulomb
potential dominates in the ultraviolet (UV) region and the linear
potential dominates in the infrared (IR) region. It motivates people
to take into account the QCD running coupling effect into the
modified metric
\cite{Csaki:2006ji,Gursoy,Zeng:2008sx,Pirner:2009gr,Brodsky:2010ur}.
In Ref.\cite{Pirner:2009gr}, Pirner and Galow have proposed a
deformed metric which resembles the QCD running coupling, and the
Pirner-Galow metric can fit the Cornell potential reasonably well.
However, as shown in Ref. \cite{Galow:2009kw} the corresponding
dilaton potential solved from the Einstein equation is unstable, and
the corresponding beta function does not agree with the QCD beta
function.

The motivation of this work \cite{He-Huang-Yan} is to show that a
deformed ${\rm AdS}_5$ metric with an explicit infrared cutoff
included in the logarithmic correction $-c_0\log[(z_{IR}-z)/z_{IR}]$
can describe the heavy quark potential as well as the QCD $\beta$
function very well, at the same time it can have a stable dilaton
potential from the gravity side.

\section{The deformed $AdS_5$ model}

To search for the possible \textit{holographic} QCD models, the most
economic way of breaking conformal invariance is to add a deformed
warp factor $h(z)$ in the metric background, and the general metric
$\mathcal {A}_s(z)$ in the string frame and in the Euclidean space
has the following form:
\begin{eqnarray}
ds^2=G_{\mu\nu}^s dX^\mu dX^\nu &=& \frac{h(z)L^2}{z^2}\left(
dt^2+d\vec{x}^2+dz^2\right) \label{h-general} \\
&=& e^{2\mathcal {A}_s(z)}\left( dt^2+d\vec{x}^2+dz^2\right).
\label{metric-general}
\end{eqnarray}

As pointed in \cite{Gursoy}, that the logarithmic term $c_0\log z$
itself cannot produce confinement, while a logarithmic correction
with an infrared cut-off in the form of $c_0\log (z_{IR}-z)$ can
have confinement at IR. Therefore, we propose the following form for
the deformed warp factor \cite{He-Huang-Yan} as
\begin{equation}
 h(z)=\exp\left( -\frac{\sigma
 z^2}{2}-c_0\ln(\frac{z_{IR}-z}{z_{IR}})\right).
 \label{metric-firstmodel}
\end{equation}
The coefficients $\sigma$ and $c_0$ can be either positive or
negative. An IR cut-off $z_{IR}$ explicitly sets in the metric,
which has the same effect as the hard-wall model. When $c_0=0$,
$\sigma>0$ and $\sigma<0$ corresponds to the soft-wall model
\cite{KKSS2006} and Andreev-Zakharov model, respectively. In
Ref.\cite{Pirner:2009gr}, in order to mimic the QCD running coupling
behavior, Pirner and Galow proposed the deformed warp factor
\begin{equation}
h_{PG}(z)=\frac{\log\left (\frac{1}{\epsilon} \right )}{\log\left
[\frac{1}{(\Lambda z)^2+\epsilon}\right ]}. \label{metric-PG}
\end{equation}
This metric with asymptotically conformal symmetry in the UV and
infrared slavery in the IR region yields a good fit to the heavy
$Q\bar Q$-potential with $\Lambda=264\,\text{MeV}$ and
$\epsilon=\Lambda^2 l_s^2=0.48$. It is worthy of mentioning that the
deformed warp factor $h_{PG}(z)$ is dominated by a quadratic term $
\sigma z^2$ in the UV regime and a logarithmic term
$-\log(z_{IR}-z)$ in the IR regime, respectively. The deformed
metric in Eq.(\ref{metric-firstmodel}) when taking the parameter of
$\sigma=0.08, c_0=1, z_{IR}=2.73 {\rm GeV}^{-1}$ can mimic the
Pirner-Galow deformed metric in Eq.(\ref{metric-PG}).

Following the standard procedure, one can derive the interquark
distance $R$ as a function of $z$
\begin{eqnarray}
R(z) &=&2 z \int_0^1 d\nu \frac{e^{2\mathcal {A}_s(z)}}{e^{2\mathcal
{A}_s(\nu z)}}\frac{1}{\sqrt{1- \left (\frac{e^{2\mathcal
{A}_s(z)}}{e^{2\mathcal {A}_s(\nu z)}}\right )^2}}.
\label{distance1}
\end{eqnarray}
The heavy quark potential can be worked out from the Nambu-Goto
string action:
\begin{eqnarray} V_{Q\bar
Q}(z)&=&\frac{1}{\pi\sigma_s}\int_0^1 d\nu e^{2\mathcal {A}_s(\nu
z)}z\frac{1}{\sqrt{1-\left ( \frac{e^{2\mathcal
{A}_s(z)}}{e^{2\mathcal {A}_s(\nu z)}}\right )^2}}.
\label{VQQ-general}
\end{eqnarray}
It is noticed that the integral in Eq.(\ref{VQQ-general}) in
principle include some poles, which induces $ V_{Q\bar
Q}(z)\rightarrow \infty$. The infinite energy should be extracted
through certain regularization procedure. The divergence of
$V_{Q\bar Q}(z)$ is related to the vacuum energy for two static
quarks.

According to the G{\"u}rsoy -Kiritsis-Nitti(GKN) framework
\cite{Gursoy}, the noncritical string background dual to the
QCD-like gauge theories can be described by the following action in
the Einstein frame:
\begin{equation}
S_{5D-Gravity}=\frac{1}{2\kappa_5^2}\int d^5x \sqrt{-G^E} \left (R
-\frac{4}{3}\partial_{\mu}\phi\partial^{\mu}\phi-V_B(\phi)\right)\,.
\label{5D_action}
\end{equation}
Where $R$ is the Ricci scalar, $\phi$ is the dilaton field, and
$V_B(\phi)$ the dilaton potential. The metric in the Einstein frame
is denoted by $G_{\mu\nu}^E$. Replacing
$A(z)=\mathcal{A}_s(z)-\frac{2}{3}\phi$, we obtain the following two
independent Einstein's equations:
\begin{eqnarray}
V_B(\phi(z))&=& -4
e^{\frac{4}{3}\phi-2\mathcal{A}_s}[(\phi')^2+3(\mathcal{A}_s')^2-4\phi'\mathcal{A}_s'
], \nonumber \\
\phi''&=& \frac{3}{2}\mathcal{A}_s'' + 2
\mathcal{A}_s'\phi'-\frac{3}{2} (\mathcal{A}_s')^2. \label{EQ-As}
\end{eqnarray}
Different from the original GKN paper, we will determine the metric
structure $\mathcal{A}_s$ from heavy quark potential, then solve the
dilaton field $\phi$ and the dilaton potential $V_B(\phi)$ from
Eq.~(\ref{EQ-As}). The resulting second order differential equation
for $\phi(z)$ needs two boundary conditions.

In the GKN framework, the scalar filed or dilaton field $\phi$
encodes the running of the Yang-Mills gauge theory's coupling
$\alpha$. For convenience, the renormalized dilaton field $\phi$ has
been defined as $\alpha=\frac{g_{YM}^2}{4 \pi}=e^{\phi}$. For a 5D
holographic model, its $\beta$ function is related to the deformed
warp factor $A(z)$ by
\begin{equation}
  \beta\,\equiv\,E\frac{d\alpha }{d
E}=\frac{e^\phi d\phi}{d A}=\frac{e^{\phi(z)}\cdot \phi'(z)}{A'(z)}.
\label{calculatebeta}
\end{equation}
The QCD $\beta$-function at 2-loop level has the following form:
\begin{equation}
  \beta(\alpha)=-b_0 \alpha^2 - b_1 \alpha^3,
\label{QCDbeta}
\end{equation}
with $b_0=\frac{1}{2\pi}(\frac{11}{3}N_c-\frac{2}{3}N_f)$, and
$b_1=\frac{1}{8\pi^2}(\frac{34}{3} N_c^2-(\frac{13}{3}
N_c-\frac{1}{N_c})N_f)$. By choosing $N_c=3$ and $N_f=4$, one has
$b_0=\frac{25}{6\pi}$ and $b_1=\frac{77}{12\pi^2}$.

\section{Heavy quark potential, QCD beta function and dilaton
potential}

We consider two cases: 1) with only quadratic correction when
$c_0=0$; 2) with only logarithmic correction when $\sigma=0$. In the
numerical calculations, we choose the ${\rm AdS}_5$ radius $L=1{\rm
GeV}^{-1}$, and the Coulomb part is fixed by choosing the string
tension $\sigma_s=0.38$.

\begin{figure}[h]
\begin{center}
\epsfxsize=6.0 cm \epsfysize=6.0 cm \epsfbox{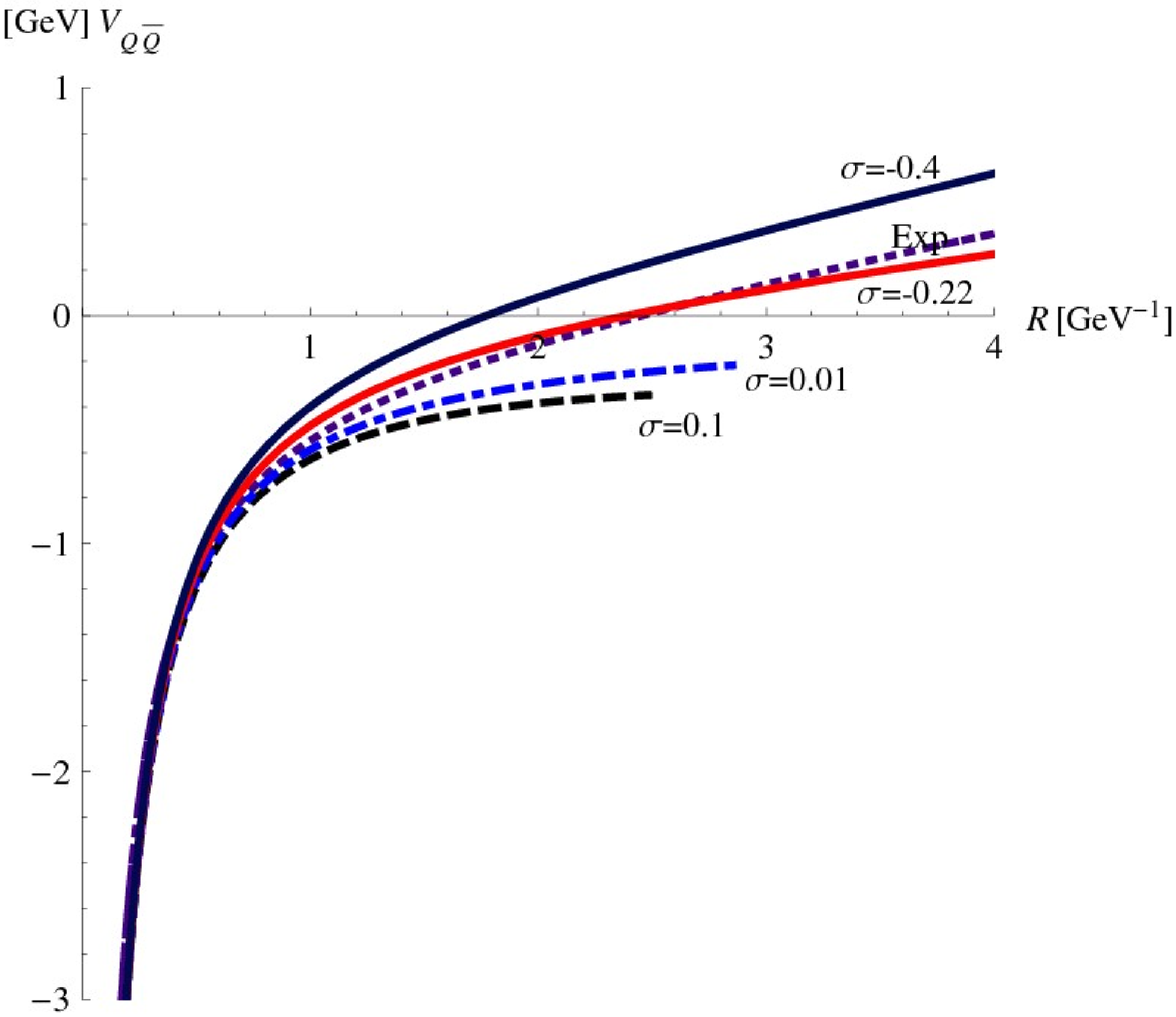}
\hspace*{0.1cm} \epsfxsize=6.0 cm \epsfysize=6.0 cm
\epsfbox{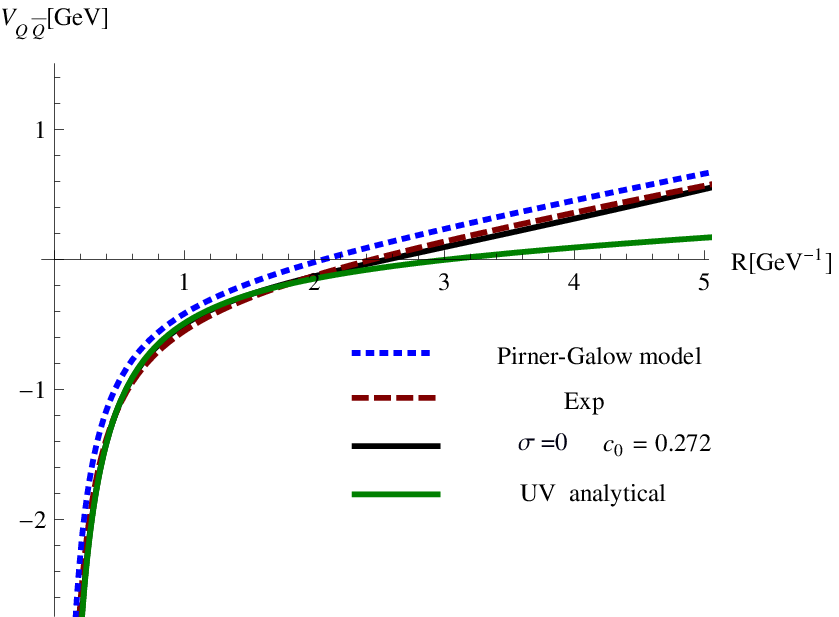} \vskip -0.05cm \hskip 0.15 cm
\textbf{( a ) } \hskip 6.5 cm \textbf{( b )} \\
\end{center}
\caption{ (a) The heavy quark potential as functions of $R$ in the
case of $c_0=0$, and $\sigma=0.1,0.01,-0.22,-0.4 {\rm GeV}^2$. (b)
The heavy quark potential as functions of the distance $R$ in the
case of $\sigma=0$ and $c_0=0.272$ and $z_{IR}=2.1 {\rm GeV}^{-1}$.
} \label{Vqq-Rz}
\end{figure}

The heavy quark potential as functions of quark anti-quark distance
$R$ for the case with only quadratic correction when $c_0=0$ is
shown in Fig. \ref{Vqq-Rz} (a), and for the case with only
logarithmic correction is shown in Fig.\ref{Vqq-Rz} (b). For the
first case, the best fit of the heavy quark potential gives
$\sigma=-0.22 {\rm GeV}^2 $, which is negative and corresponds to
the Andreev-Zakharov model. For the case with only logarithmic
correction when $\sigma=0$, the best fitted heavy quark potential
(the black solid line in Fig.\ref{Vqq-Rz} (b)) gives $c_0=0.272 {\rm
GeV}^2$ and $z_{IR}=2.1 {\rm GeV}^{-1}$. The results are compared
with that from the Pirner-Galow model (short dashed line) and the
experimental data (the long dashed line) and the UV analytical
result in Fig.\ref{Vqq-Rz} (b).

\begin{figure}[h]
\begin{center}
\epsfxsize=6.0 cm \epsfysize=6.0 cm \epsfbox{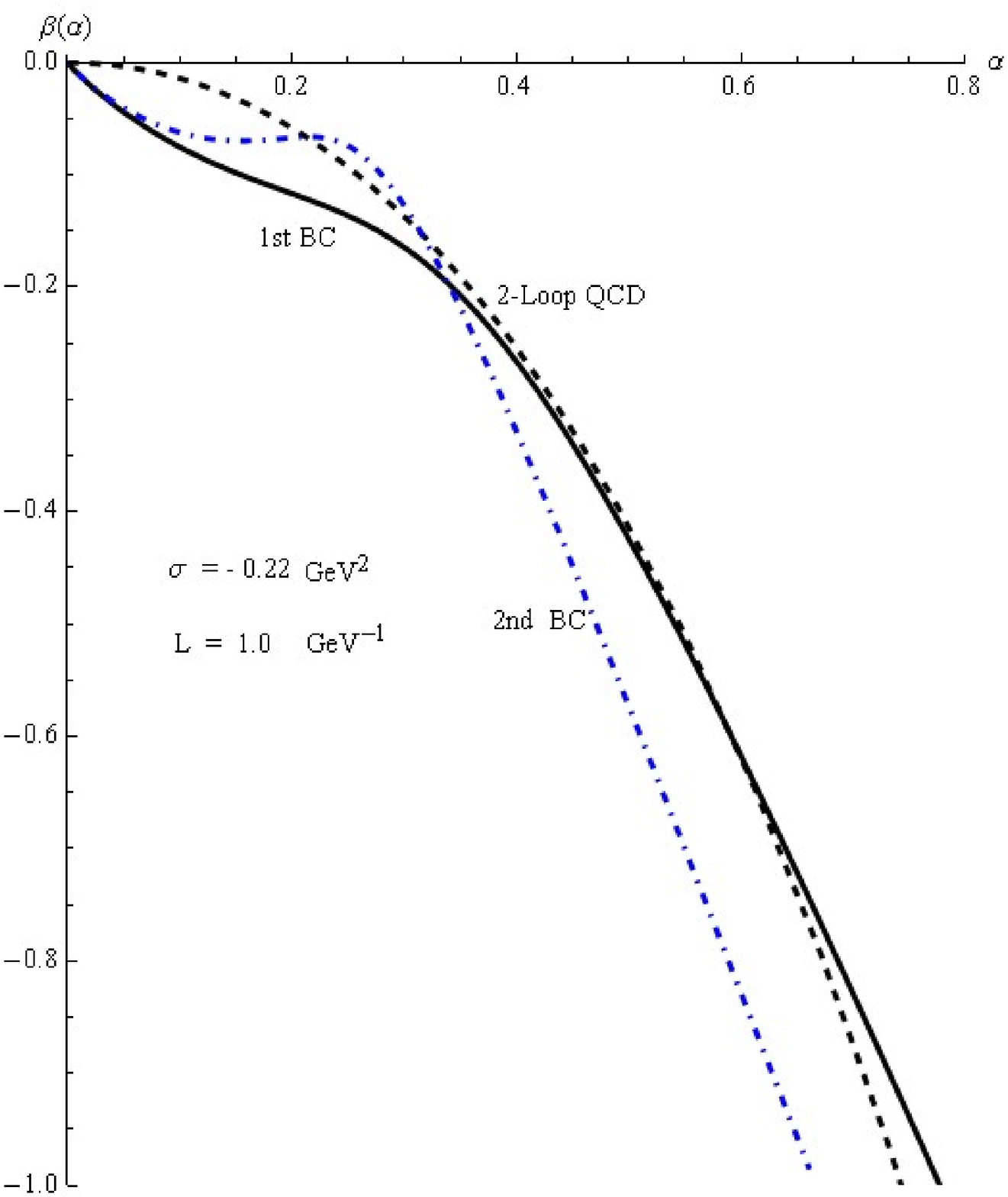}
\hspace*{0.1cm} \epsfxsize=6.0 cm \epsfysize=6.0 cm
\epsfbox{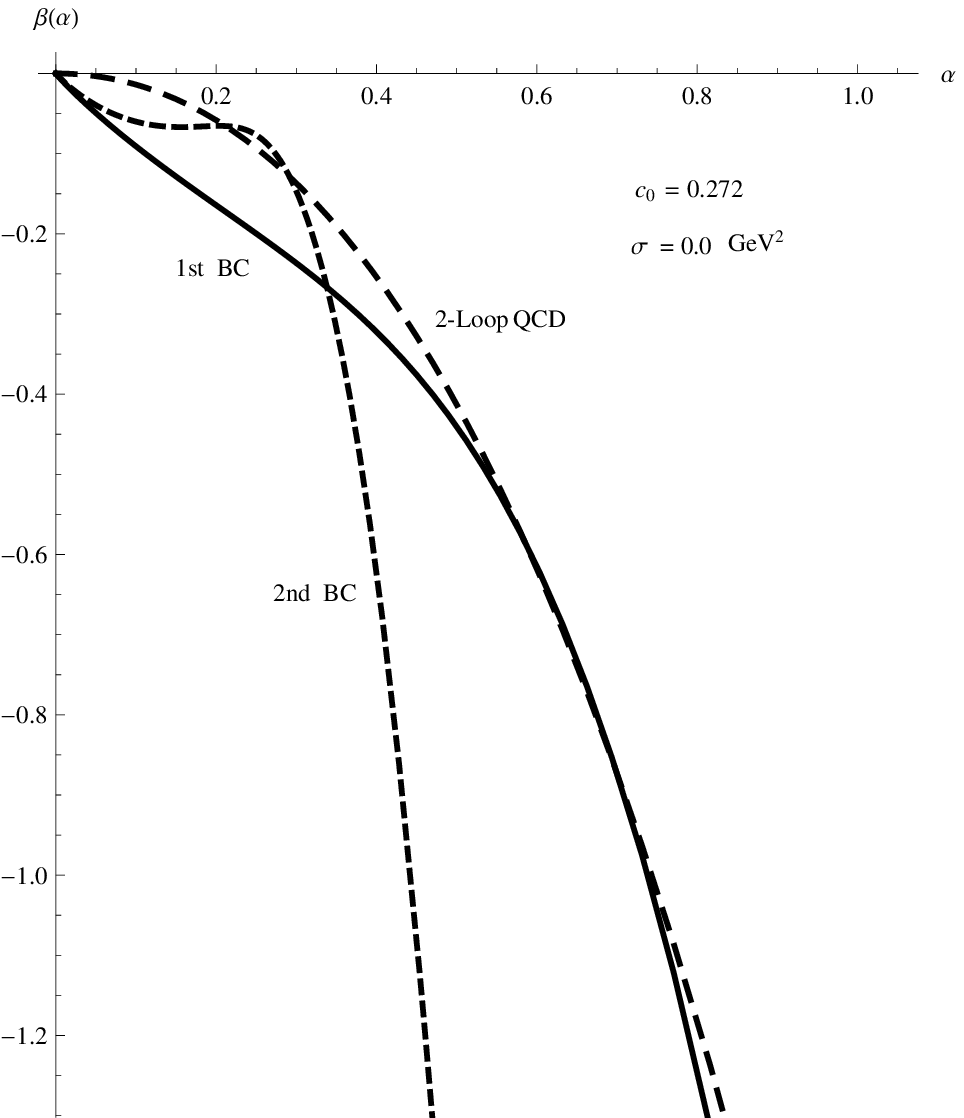} \vskip -0.05cm \hskip 0.15 cm
\textbf{( a ) } \hskip 6.5 cm \textbf{( b )} \\
\epsfxsize=6.0 cm \epsfysize=6.0 cm \epsfbox{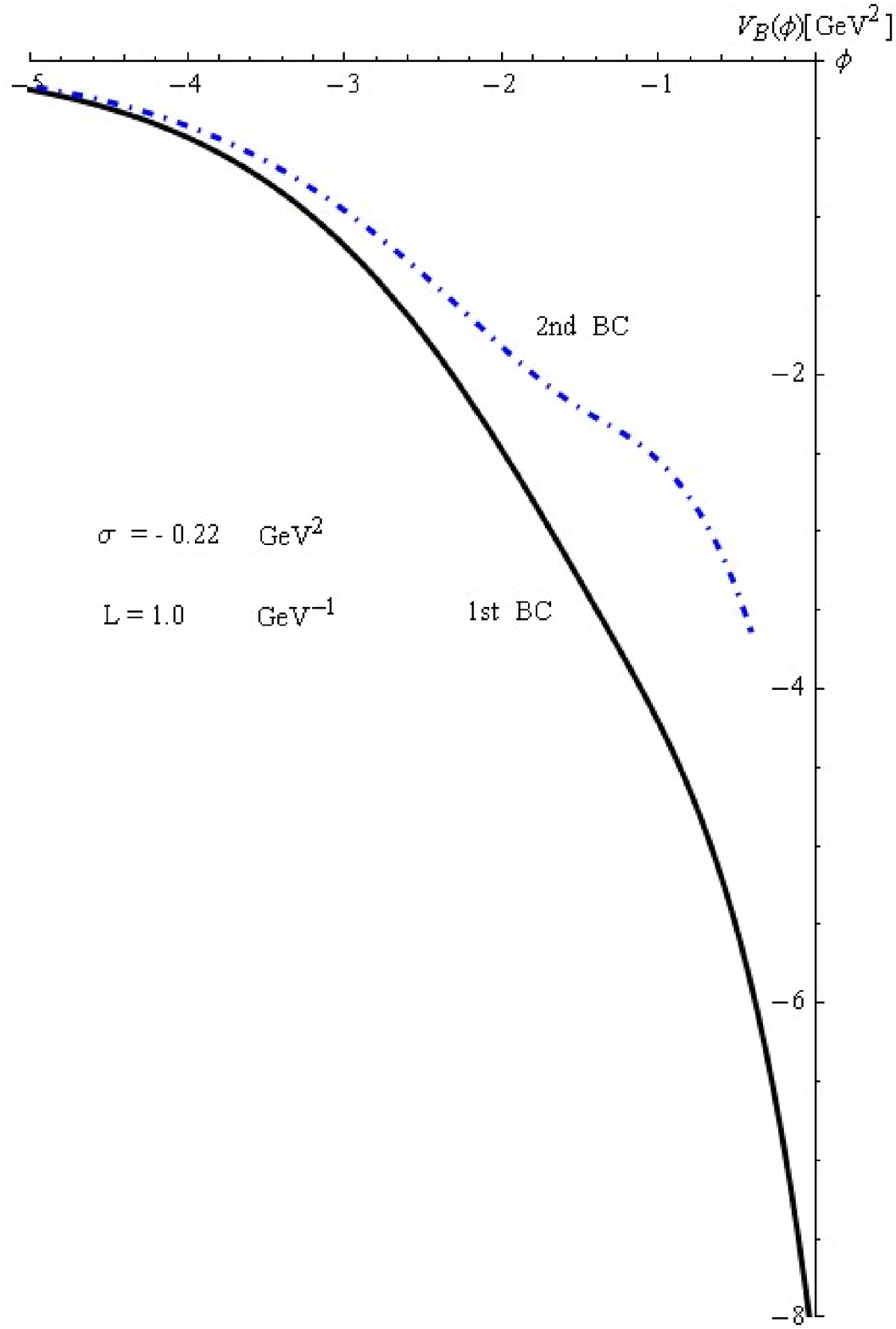}
\hspace*{0.1cm} \epsfxsize=6.0 cm \epsfysize=6.0 cm
\epsfbox{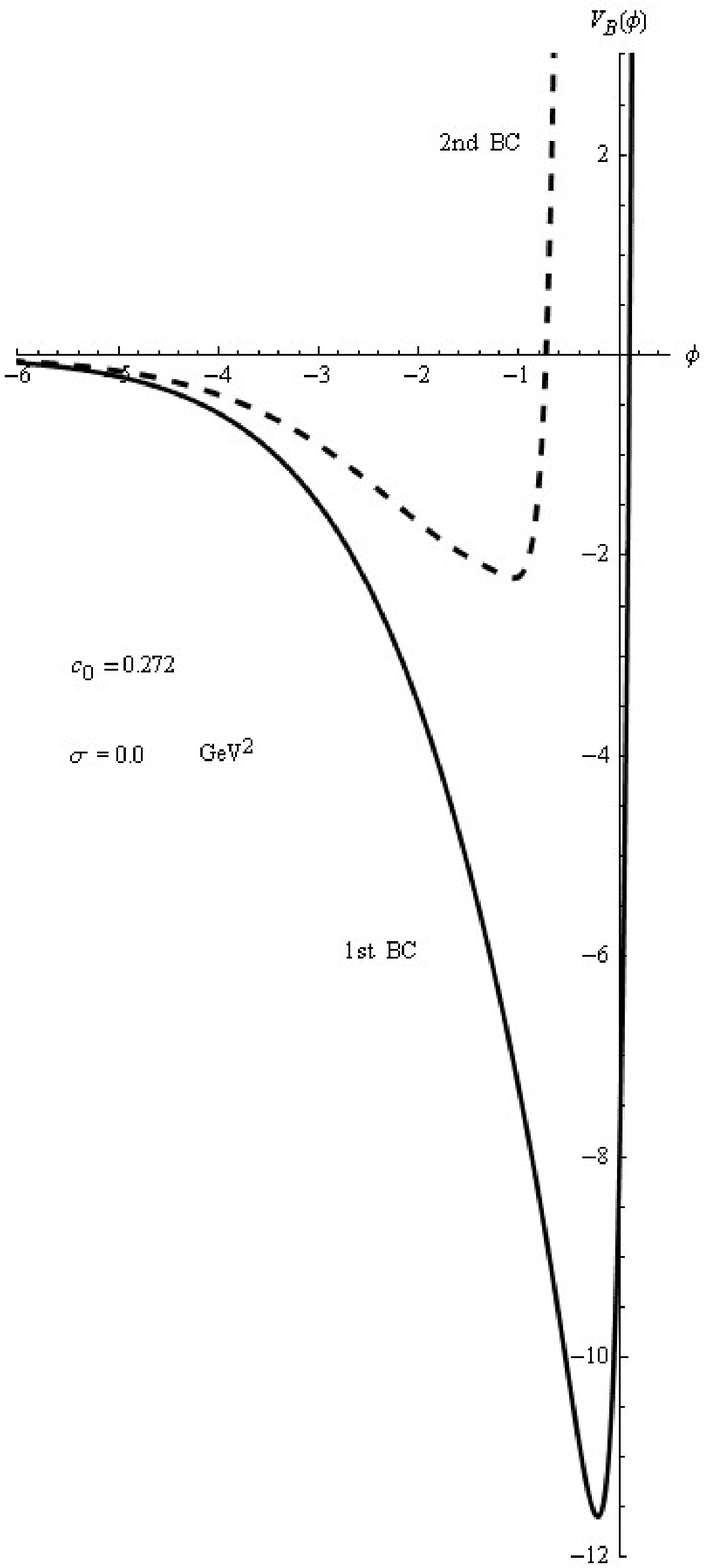} \vskip -0.05cm \hskip 0.15 cm
\textbf{( c ) } \hskip 6.5 cm \textbf{( d )} \\
\end{center}
\caption[]{The beta function as a function of coupling constant
$\alpha$ and the dilaton potential as a function of $\phi$ in the
case with only quadratic correction ($c_0=0$, $\sigma=-0.22$) and in
the case with only logarithmic correction ($\sigma=0$, $c_0=0.272$
and $z_{IR}=2.1 {\rm GeV}^{-1}$). The boundary conditions are
described in Eqs.(\ref{BC-c0zero}) and (\ref{BC-sigmazero}).}
 \label{beta-Vphi}
\end{figure}

The $\beta$ function as a function of $\alpha$ and the dilaton
potential as a function of $\phi$ are shown in Fig. \ref{beta-Vphi}
for the case of quadratic correction and for the case of logarithmic
correction, respectively. For the case with only quadratic
correction, the used two types of boundary conditions are:
\begin{eqnarray}
& & {\rm 1stBC}: \phi(z=0.87)={\rm log}(0.25), ~
\phi'(z=0.87)=0.9, \nonumber \\
& & {\rm 2ndBC}: \phi(z=0.87)={\rm log}(0.25), ~ \phi(z=0.38)={\rm
log}(0.18). \label{BC-c0zero}
\end{eqnarray}
For the case with only logarithmic correction, the used two types of
boundary conditions are:
\begin{eqnarray}
& & {\rm 1st BC}:  \phi(z=0.9)={\rm log}(0.25), ~
\phi'(z=0.9)=1.7, \nonumber \\
& & {\rm 2nd BC}:  \phi(z=0.9)={\rm log}(0.25), ~ \phi(z=0.39)={\rm
log}(0.185). \label{BC-sigmazero}
\end{eqnarray}

For both cases with only quadratic correction and with only
logarithmic correction, it is found that by using the second type
boundary condition, i.e, the boundary condition used in
\cite{Galow:2009kw}, the produced $\beta$ function is not a
monotonic function of coupling $\alpha$. This behavior is due to the
fixing running coupling constant at two points. By using the first
type of boundary condition, the produced $\beta$ function is
monotonically decreasing with the coupling constant $\alpha$, and it
agrees reasonably well with the QCD $\beta$ function at 2-loop
level, which is shown by dashed line in Fig. \ref{beta-Vphi}.

For the case with only quadratic correction, it is found that for
both types of boundary conditions, $V_B(\phi)$ decreases with
$\phi$, the dilaton potential in the IR regime is not bounded from
below, which might indicate an unstable vacuum. This behavior is
also shown in the Pirner-Galow model in Ref. \cite{Galow:2009kw}.
However, in the model with only logarithmic correction, it is found
that for both types of boundary condition, the dilaton potential
$V_B(\phi)$ is stable which is bounded from below in the IR.

\section{Conclusion}

We found that in the deformed $AdS_5$ model with only logarithmic
correction in the deformed warp factor, the heavy quark potential
can be fitted very well and the beta function of the running
coupling agrees well with QCD $\beta$ function at 2-loop level.
Comparing with the Andreev-Zakharov model and the Pirner-Galow
model, the corresponding dual dilaton potential is stable in the
model with only logarithmic correction.

\section*{Acknowledgements}
M.H. thanks the Yukawa Institute for Theoretical Physics at Kyoto
University and the organizers of NFQCD2010 for their hospitality.
The work of M.H. is supported by CAS program "Outstanding young
scientists abroad brought-in", CAS key project KJCX3-SYW-N2,
NSFC10735040, NSFC10875134, and K.C.Wong Education Foundation, Hong
Kong.

\end{document}